\documentclass{article} 
\usepackage{preprint, times}
\usepackage{amsmath}
\usepackage{graphicx}
\usepackage{url}
\usepackage{color}
\usepackage{listings}
\usepackage{pgfplots}

\lstdefinelanguage{scala}{
  keywords    = {def, val, let, return, new, if, then, else, while, for, throw, =},
  morecomment = [l]{//},
  morecomment = [s]{/*}{*/},
  morestring  = [b]',
  morestring  = [b]",
  sensitive   = true,
}

\definecolor{codegreen}{rgb}{0, 0.6, 0}
\definecolor{codegray}{rgb}{0.5, 0.5, 0.5}
\definecolor{codepurple}{rgb}{0.58, 0, 0.82}
\definecolor{backcolour}{rgb}{0.99, 0.99, 0.99}

\lstdefinestyle{base}{
  language=Scala,
  basicstyle=\tt\scriptsize,
  backgroundcolor=\color{backcolour},
  commentstyle=\ttfamily\color{codegreen},
  keywordstyle=\bfseries\color{magenta},
  numberstyle=\tiny\color{codegray},
  stringstyle=\ttfamily\color{codepurple},
  showstringspaces=false,
  numbers=none,
  numbersep=5pt,
  captionpos=b,
  literate={~}{$\sim$}{1},
  escapeinside={(*@}{@*)},
}

\title{DeepDSL: A Compilation-based Domain-Specific Language for Deep Learning}

\author{Tian Zhao \& Xiao Bing Huang \\
Department of Computer Science\\
University of Wisconsin -- Milwaukee\\
Milwaukee, WI, USA \\
\texttt{\{tzhao,xiaobing\}@uwm.edu} \\
\And
Yu Cao \\
Department of Computational Neuroscience\\
The University of Massachusetts, Lowell\\
Lowell, MA, USA\\
\texttt{ycao@cs.uml.edu}
}

\begin{document}

\maketitle

\begin{abstract}
In recent years, \textit {Deep Learning} (DL) has found great success in domains such as multimedia understanding. However, the complex nature of multimedia data makes it difficult to develop DL-based software. The state-of-the-art tools, such as Caffe, TensorFlow, Torch7, and CNTK, while are successful in their applicable domains, are programming libraries with 
fixed user interface, internal representation, and execution environment. This makes it difficult to implement portable and customized DL applications.

In this paper, we present {\it DeepDSL}, a \textit {domain specific language} (DSL) embedded in Scala, that compiles deep networks written in DeepDSL to Java source code. Deep DSL provides 
(1) intuitive constructs to support compact encoding of deep networks; 
(2) symbolic gradient derivation of the networks; 
(3) static analysis for memory consumption and error detection; and 
(4) DSL-level optimization to improve memory and runtime efficiency. 

DeepDSL programs are compiled into compact, efficient, customizable, and portable Java source code, which operates the CUDA and CUDNN interfaces running on Nvidia GPU
via a \textit {Java Native Interface} (JNI) library. 
We evaluated DeepDSL with a number of popular DL networks. 
Our experiments show that the compiled programs have very competitive runtime performance and memory efficiency compared to the existing libraries.
\end{abstract}

\section{Introduction}
\label{intro}
Multimedia is increasingly becoming the "biggest big data" as the most important and valuable source for insights and 
information \cite{chen2015guest}. 
Recently, a new set of machine learning algorithms named "Deep Learning" (DL) \cite{DeepLearning2015LecunBengioHinton}, 
which aims at learning multiple levels of representation and abstraction that help infer knowledge from multimedia data (e.g. text, image, audio, and video) is making astonishing gains in machine vision, speech recognition, 
multimedia analysis, and drug designing. 

However, current tools, such as Theano~\cite{bergstra+al:2010-scipy}, Torch7~\cite{collobert:2011c}, Caffe~\cite{jia2014caffe}, \textit {Computational Network Toolkit} (CNTK)~\cite{export:226641}, and TensorFlow~\cite{2016arXiv160304467A}, while are efficient in 
their applicable domains, are essentially application libraries with some inherent limitations. 

As with all programming libraries, the DL libraries have fixed bindings for
key data structures such as tensors and tensor related computations.
Users have to adhere to the data structure, which limits their ability to 
apply application-specific optimization or port it to target runtime platforms. 
The internal representation of their control flow logic is opaque to users. For example, 
TensorFlow and CNTK use directed acyclic graphs to represent the DL network computation and 
generate runtime binaries from the graphs. 
However, these graphs are not designed for user-level access, which
limits the runtime platforms of the DL applications to what the libraries provide.

In general, the current libraries have to be built to specific platforms that they are designed for, 
which can be difficult for platforms such as Windows. Also, changing the implementation of specific
type of layers or data structure is very challenging without in depth understanding of the underlying
implementation. This limits their portability and reusability.

To address these limitations, we present DeepDSL, a domain specific language embedded in Scala, for developing 
DL applications. DeepDSL allows users to define DL networks as tensor functions. 
Unlike the existing DL libraries, DSL tensors are not built-in entities. Instead, they are
defined as indexed scalar expressions. This exposes tensor related computation at DSL level. 
As a result, the symbolic gradient derivation of the DL network is fully abstract and the resulting DSL program
allows compiler-based optimizations such as code motion and common sub-expression elimination.

DeepDSL compiler translates the optimized DSL program into a Java source program that
is compact, efficient, customizable, and portable.
The generated Java source only depends on a small Java library that calls CUDA through a JNI 
library called JCuda~\footnote{http://www.jcuda.org}. 
Since JVM is supported on all major operating systems, the generated Java source
can run anywhere with CUDA. Also, since the generated Java source is compact and human readable, 
users can customize it easily through an editor or IDE such as eclipse~\footnote{http://www.eclipse.org}. The generated Java source automatically saves the learned parameters into files after a training period is over. When user starts the program again (perhaps after adjusting some parameters such as momentum and learning rate), it automatically loads the saved parameters and continues the training from where
it stopped at the previous execution. The code also supports loading parameters trained with different data for fine tuning purpose.

DeepDSL is able to statically analyze the DSL program to detect network design errors such as mismatching tensor dimensions
before compiling the DSL program into Java source.
It statically analyzes the memory consumption used at each step of the computation and produces a table detailing
the memory usage that would occur at runtime, which includes the memory for feature maps, gradient maps, parameter weights,
and convolution workspace. 
It also uses the static information to reschedule computation so that tensor memory can be freed as early as possible to 
reduce memory consumption at runtime. Such processing has demonstrated to have great benefit. For example, DeepDSL continues to run well under the GPU memory limit on the testing server with a single GPU when the batch size of ResNet is increased from 32 to 64, while both Caffe and Tensorflow fail due to out of memory exception.

DeepDSL is available at \url{https://github.com/deepdsl/deepdsl}.

The rest of the paper is organized as follows. We give an overview of DeepDSL in Section~\ref{overview} and explain the DSL syntax using examples in Section~\ref{syntax}. We discuss the intermediate representation in Section~\ref{inter} and code generation in Section~\ref{comp}. We present details of performance evaluation using DeepDSL in Section~\ref{perform} and related work in Section~\ref{rel}. We conclude in Section~\ref{conclude}. 

\section{Overview}
\label{overview}

\begin{figure}[t]
\begin{center} \includegraphics[scale=0.4]{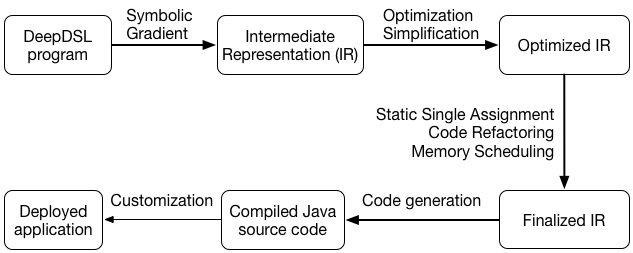} \end{center}
\caption{Basic workflow of DeepDSL.} 
\label{fig:overview}
\end{figure}

DeepDSL directly encodes the mathematical representation of DL networks, where each layer
is represented as a tensor function. The entire network is then represented as a composition of these functions.
DeepDSL symbolically derives the partial derivatives of the tensor functions with respect to tensor variables
so that the backward gradients of network parameters are generated automatically. 

A high-level overview of DeepDSL is shown in Figure~\ref{fig:overview}. A DeepDSL program is compiled in several stages. At the first stage, the backward gradients of deep networks are derived symbolically to become the intermediate representation (IR). The IR expressions are in turn passed through a series of simplification and optimization at the second stage. At the third stage, DeepDSL compiler performs a SSA (Static Single Assignment) transformation of the optimized IR to break down complex expressions. Redundant computation is eliminated at this stage and the resulting expressions are reordered to optimize memory usage. Memory deallocation and in-place computation are also scheduled at this stage. Lastly, the finalized IR expressions are translated to Java source code.

DeepDSL supports two mode of computation: memory efficient or runtime efficient. In the memory efficient mode, tensor memory in GPU will be dynamically allocated and deallocated, which might decrease runtime performance. In the runtime efficient mode, tensor memory in GPU is reused and not deallocated until the end of the training. In this mode, more memory may be used but with greater runtime performance. To make the switch, the user only needs to switch a flag to the generated Java source.
The memory efficient mode can be used for machines with limited GPU memory. Further memory reduction can be achieved by placing a limit on the (convolution) workspace memory.

\section{Syntax}
\label{syntax}

\begin{figure}[t]
\begin{lstlisting}[language=Scala,numbers=left,xleftmargin=4mm]
val K = 10 // # of classes 
val N = 500; val C = 1; val N1 = 28; val N2 = 28 // batch size, channel, and x/y size
 
// Specifying training (and test) dataSet
val y = Vec._new(Mnist, "label", "Y", N)            // labels
val x = Vec._new(Mnist, "image", "X", N, C, N1, N2) // images
   
val cv1 = CudaLayer.convolv("cv1", 5, 20)       // kernel size (5,5), output channel 20
val cv2 = CudaLayer.convolv("cv2", 5, 50)
val mp = CudaLayer.max_pool(2)                  // max pooling, kernel 2 stride 2
val flat = Layer.flatten(4, 1)                  // flatten a 4-D tensor from axis 1 to 3      
val f = Layer.full("fc1", 500)                  // fully connected layer, output 500
val f2 = Layer.full("fc2", K)
val relu = CudaLayer.relu(2)                    // 2-D ReLU activation
val softmax = CudaLayer.softmax                 // softmax 
      
// o is a left-associative operator for function composition 
val network = f2 o relu o f o flat o mp o cv2 o mp o cv1 

val x1 = x.asCuda                               // load x to GPU
val y1 = y.asIndicator(K).asCuda                // turn each label into an indicator vector 
val c = (Layer.log_loss(y1) o softmax o network) (x1) // training loss
val p = (Layer.precision(y1) o network) (x1)          // test accuracy

val param = c.freeVar.toList                    // parameters to be trained

// output file, train and test iteration, learn rate, momentum, decay, gradient cropping (0 means none)
val solver = Train("lenet", 1000, 10, 0.01f, 0.9f, 0.0005f, 0)  

val loop = Loop(c, p, (x, y), param, solver)    // training and testing loop 
cudnn_gen.print(loop)                           // generate Java source program
\end{lstlisting}
\caption{DeepDSL code for training and testing {\em Lenet}.}
\label{fig:lenet}
\end{figure}
Figure~\ref{fig:lenet} shows the complete implementation for compiling a program to train and test
{\em Lenet}~\cite{lecun1998gradient}.
Since DeepDSL is embedded in Scala, the program is in Scala syntax and it can be
compiled and run with a programming tool such as eclipse. 
This program consists of variable declarations of the form \verb|val x = e|, where \verb|val|
starts a declaration for the variable \verb|x| and assigns it with the value of \verb|e|.

Line 5 and 6 declare the tensors that represent labels and images for the training data. 
We also use the same variables for testing since the DSL compiles the same variables into different
code for training and testing.

Line 8--15 declare the tensor functions that represent the layers in the network. Most of the layers
are self-explanatory except \verb|val flat = Layer.flatten(4, 1)|, which is used to convert the 4-D
tensor returned by the last pooling layer into a 2-D layer for the next fully connected layer.

Line 18 constructs the network as function compositions using the operator \verb|o|, which is
left associative. For example, \verb|f2 o relu o f| should be read as
\verb|(f2 o relu) o f|. A composed function such as \verb|network| is still a function. 

Line 22 defines the expression that represents the loss of the network when applied to the training data.
Line 23 defines the testing accuracy of the trained network.

Line 25 extracts the parameters such as weights and biases from the loss expression.
Line 28--31 defines the solver object, passes it to the loop object for training and testing,
and then generates the Java source code.

\paragraph{Layer reuse}
Since each layer is a tensor function, for the layers such as ReLU and pooling that do not contain
parameters, we can simply reuse them in a network.
For example, in the following definition for {\em Alexnet}, \verb|relu2| (2 dimensional),
\verb|relu| (4 dimensional), \verb|pool| (max pooling), \verb|drop| (drop out), 
and \verb|lrn| (local response normalization) are reused.

\begin{lstlisting}[language=Scala]
    val network =       full8 o 
         drop o relu2 o full7 o 
         drop o relu2 o full6 o flat o 
          pool o relu o cv5 o  
                 relu o cv4 o 
                 relu o cv3 o 
    pool o lrn o relu o cv2 o  
    pool o lrn o relu o cv1 
\end{lstlisting}

Layer function reuse simplifies the definitions of deep networks. For {\em Alexnet}, only 5 convolution layers
and 3 fully connected layers need to be defined separately.
Note that the above definition can be written in just one line and the line breaks are only for clarity.

\paragraph{Network reuse}
For complex network such as {\em Googlenet}, we can define reusable subnet to achieve compact definitions.
For example, the Scala method \verb|inception| below returns a tensor function that
represents an inception subnet in {\em Googlenet}. 
\begin{lstlisting}[language=Scala,numbers=left,xleftmargin=4mm]
val w = Param.xavier                    // Xavier initialization for weight
val b0 = Param.const(0, 2, 0)           // constant 0 for bias, learn rate/decay multiplier 2 and 0 
val b02 = Param.const(0.2f, 2, 0)       // constant 0.2 for bias
val ipool = CudaLayer.max_pool(3, 1, 1) // max pooling kernel size, stride, and padding

def inception(n: Int) = {
  // convolution name, kernel size, channel, stride, padding, weight and bias configuration 
  val icv1 = CudaLayer.convolv(s"cv${n}1", 1, 64, 1, 0, w, b02) 
  val icv2 = CudaLayer.convolv(s"cv${n}2", 1, 96, 1, 0, w, b02)
  val icv3 = CudaLayer.convolv(s"cv${n}3", 3, 128, 1, 1, w, b02)
  val icv4 = CudaLayer.convolv(s"cv${n}4", 1, 16, 1, 0, w, b02)
  val icv5 = CudaLayer.convolv(s"cv${n}5", 5, 32, 1, 2, w, b02)
  val icv6 = CudaLayer.convolv(s"cv${n}6", 1, 32, 1, 0, w, b02)
  
  val p = Vec._new(4)                             // a 4-dimensional tensor variable

  // a tensor function with parameter p 
  VecFun(p, CudaLayer.concat( (relu o icv1)(p),   // concatenation of 4 subnets connected to p  
                (relu o icv3 o relu o icv2)(p),  
                (relu o icv5 o relu o icv4)(p),
                      (relu o icv6 o ipool)(p) )
  )
}
\end{lstlisting}
Using the \verb|inception| method, we can define three subnets 
that are used to define the test accuracy \verb|p| (line 6 below) of the main branch of {\em Googlenet}. 
\begin{lstlisting}[language=Scala,numbers=left,xleftmargin=4mm]
val network3 = full7 o flat o drop o pool7 o inception(9) o inception(8) o pool o inception(7) 
val network2 = inception(6) o inception(5) o inception(4) 
val network1 = inception(3) o pool o inception(2) o inception(1) o 
                              pool o lrn o relu o cv3 o relu o cv2 o lrn o pool o relu o cv1 

val p = Layer.precision(y1)(network3(network2(network1(x1))))  // accuracy at main branch
\end{lstlisting}
The three subnets are also used to define the training loss \verb|c| (line 16 below) 
that adds up the losses of the three branches of {\em Googlenet}.
\begin{lstlisting}[language=Scala,numbers=left,xleftmargin=4mm]
def branch(n: Int) = {    // a subnet reused in the two side branches of Googlenet
  val cv = CudaLayer.convolv(s"b${n}cv", 1, 128, 1, 0, w, b02)
  val f1 = Layer.full(s"b${n}fc1", 1024, w, b02)
  val f2 = Layer.full(s"b${n}fc2", K, w, b0)

  f2 o drop2 o relu2 o f1 o flat o relu o cv o bpool
}
val stage2 = {            // Vec2ScalarFun defines a function from tensor to scalar
  val p = Vec._new(4)
  Vec2ScalarFun(p, softmax_loss(network3(p)) + softmax_loss(branch(2)(p)) * Real(0.3f, "loss2"))
}
val stage1 = {            // Real(0.3f, "loss1") is a named constant of value 0.3
  val p = Vec._new(4)           
  Vec2ScalarFun(p, stage2(network2(p)) + softmax_loss(branch(1)(p)) * Real(0.3f, "loss1"))
}
     
val c = (stage1 o network1)(x1)  // training loss of the three branches
\end{lstlisting}
Other than some definitions of shared layers such as activation, pooling, normalization, drop out, and
softmax loss, this is the complete definition of {\em Googlenet}. 

This compact style of definition is similar to that of Theano, Tensorflow, Torch, and Mxnet.
In the example, we used two types of functions \verb|VecFun| and \verb|Vec2ScalarFun|, which
model computation that takes a tensor as input and returns a tensor or scalar respectively.
These functions can be composed or applied to arguments.
When applied, they are similar to functions in Theano, Tensorflow, and Mxnet. 
When composed, they are similar to the sequential container of Torch.

\section{Intermediate Representation}
\label{inter}

The unique advantage of DeepDSL is that it is entirely high-level so that 
it permits static analysis of the deep networks for error checking, memory analysis,
optimization, and code generation. 

While DeepDSL compiler is implemented in Scala, it has no runtime dependency on code in Scala at all. 
The whole purpose of using Scala as the host language for DeepDSL is that Scala is a strongly typed language with
flexible syntax. As a result, the syntax of DeepDSL can resemble that of a standalone DSL without having a parser.  
After taking symbolic gradients, a DeepDSL program is immediately evaluated to intermediate representation (IR), 
which is essentially an abstract syntax tree (AST). 
DeepDSL compiler analyzes its IR expressions by performing a series of optimization and simplification steps. 
During this process, DeepDSL checks the compatibility of the layers, infers concrete dimensions for variables,
removes duplicated computation, and optimizes IR expressions for code generation. 

\begin{figure}[t]
\begin{lstlisting}[language=Scala]
IR expression                               Dimensions    Current mem    Total     w/o dealloc
---------------------------------------------------------------------------------------------
val X7 = Cuda(X)                              500 1 28 28    1.568000    1.568000    1.568000
val X8 = Convolv(1,0)(X7,cv1_W,cv1_B)         500 20 24 24  23.040001   24.608000   24.608000
val X9 = Pooling(2,2,0,true)(X8)              500 20 12 12   5.760000   30.368000   30.368000
val X10 = Convolv(1,0)(X9,cv2_W,cv2_B)        500 50 8 8     6.400000   36.768002   36.768002
val X11 = Pooling(2,2,0,true)(X10)            500 50 4 4     1.600000   38.368000   38.368000
val X12 = (X11[1><3])(i | @) * (fc1_W)(j | @) 500 500        1.000000   39.368000   39.368000
val X14 = (X12 + (i) => fc1_B)                500 500        0.000000   39.368000   39.368000
val X15 = ReLU()(X14)                         500 500        0.000000   39.368000   39.368000
val X16 = (X15)(i | @) * (fc2_W)(j | @)       500 10         0.020000   39.388000   39.388000
val X18 = (X16 + (i) => fc2_B)                500 10         0.000000   39.388000   39.388000
val X19 = Softmax()(X18)                      500 10         0.020000   39.408001   39.408001
Dealloc(X18)                                                -0.020000   39.388000   39.408001
val X20 = Cuda(Indicator(Y, 10))              500 10         0.020000   39.408001   39.408001
val X21 = Log X19.copy                        500 10         0.020000   39.428001   39.428001
val X52 = 1/(X19.copy)                        500 10         0.020000   39.448002   39.448002
Print(((0 - (X20 . X21)) / |500|))                           0.000000   39.448002   39.448002

................... 30 lines omitted .......................................................

cv2_B <~~ X71 * d_Convolv(1,0)()/d_cv2_B                     0.000000   36.768002   48.448002
val X72 = X71 * d_Convolv(1,0)(cv2_W)/d_X9    500 20 12 12   5.760000   42.528000   54.208000
cv2_W <~~ X71 * d_Convolv(1,0)(X9)/d_cv2_W                   0.000000   42.528000   54.208000
Dealloc(X71)                                                -6.400000   36.127998   54.208000
val X74 = X72 * d_Pooling(2,2,0,true)(X9,X8)/d_X8         
                                              500 20 24 24  23.040001   59.167999   77.248001
Dealloc(X72)                                                -5.760000   53.408001   77.248001
Dealloc(X9)                                                 -5.760000   47.647999   77.248001
Dealloc(X8)                                                -23.040001   24.608000   77.248001
cv1_B <~~ X74 * d_Convolv(1,0)()/d_cv1_B                     0.000000   24.608000   77.248001
cv1_W <~~ X74 * d_Convolv(1,0)(X7)/d_cv1_W                   0.000000   24.608000   77.248001
Dealloc(X74)                                               -23.040001    1.568000   77.248001
Dealloc(X7)                                                 -1.568000    0.000000   77.248001
\end{lstlisting}
\caption{A portion of the IR expressions and memory information compiled from Lenet}
\label{fig:ir}
\end{figure}

The IR expressions of DeepDSL are also abstract and human readable.
For example, Figure~\ref{fig:ir} shows a portion of the IR expressions for Lenet, where the first column shows 
an IR expression that represents a single-step computation, 
the second column shows the dimensions of the tensor being computed if applicable,
the third column shows the memory usage of that tensor,
the fourth column shows the current memory consumption if memory is dynamically allocated and deallocated,
and the last column shows the memory consumption if memory is reused instead of deallocated.

IR expression such as Line 14 is for GPU memory deallocation.
DeepDSL compiler analyzes the dependencies of the IR expressions, reorders them, and determines the earliest
point where a tensor can be freed. For example, the last use of the tensor \verb|X18| in at line 13, it can be freed next.
The tensor \verb|X8| cannot be freed until much later since it is used at line 26.

If we compile IR expressions such as line 14 to actual memory deallocation, then the maximum dynamic memory
consumed is peaked at line 27, which is about 59 MB. However, frequent memory allocation and deallocation in Nvidia GPU reduces runtime performance. Therefore, DeepDSL runtime library (implemented in Java) supports memory reuse instead of deallocation. DeepDSL runtime maintains a pool of allocated memory blocks and when a tensor is freed, its memory is returned to the pool and when a tensor is allocated, the runtime tries to find a suitable block in the pool first. With memory reuse, the memory consumption always peaks at the last line, which is about 77 MB. Note that the above memory figure is for storing intermediate results such as gradients; the static memory allocated for parameters and convolution workspace are calculated separately.

DeepDSL compiler generates Java source code for each of the IR expressions. For example, line 3 loads a batch of images into GPU memory. 
Line 4 and line 5 perform forward convolution and pooling computation respectively. Line 18 prints out the training loss. Line 22 updates of the bias of the second convolution layer with its gradient. 

Some computation (e.g.~\verb|Log|) is always in-place. Therefore we make a copy of a tensor if it is passed to such computation (e.g. \verb|Log X19.copy|). Gradient update such as \verb|cv1_W <~~ X74 * d_Convolv(1,0)(X7)/d_cv1_W| may be implemented as in-place computation as well by directly updating the tensor \verb|cv1_W| when computing the backward filter gradient of the convolution layer \verb|cv1|. 

\section{Compilation}
\label{comp}

A DeepDSL program compiles to a Java source program, which uses a small Java library 
to call CUDA and CuDNN via a JNI wrapper. 
The compiled Java code does not depend on DeepDSL compiler or Scala, which makes it more portable
and easier to integrate in other applications.

Most of the current tools use platform dependent programming languages such as C, Python, 
and Lua, which compiles to specific binary for each installation. 
Since our compiled program is Java, it runs directly on any platforms that support JVM. 
Compilation of Java is trivial for most computing platforms. 
For example, the Java source generated by DeepDSL on a Windows laptop can run on a Linux server 
without any change. While it takes efforts to install tools like Tensorflow, Caffe, or Torch on machines 
with different system architectures, running the Java code generated from DeepDSL require very little work.

\paragraph{Gradient Derivation and Optimization}

The gradient derivation and optimization are implemented by the \verb|Loop| class called in code below:
\begin{lstlisting}[language=Scala]
val loop = Loop(loss, accuracy, (x, y), param, solver)
\end{lstlisting}

To derive the gradient of a scalar expression \verb|loss| with respect to a tensor variable \verb|p|, 
we can write \verb|val grad = loss.grad(p)|, which evaluates to a tensor expression.
The gradient updates are formed by expressions of the form
${\tt Update}({\tt p}, {\tt grad}, \alpha, \beta)$, which represents the computation ${\tt p} = \beta * {\tt p} + \alpha * {\tt grad}$.

The gradient updates of all parameters together with the loss expression
are then passed to optimization functions to obtain a list of IR expressions ready for code generation.
The optimization functions implement simplification, loop merging, code motion, vectorization, SSA transformation, 
common sub-expression elimination, inlining, tensor deallocation, and code scheduling.

\paragraph{Generated code}
The compiled Java code includes just one class. 
The class fields include the objects that handle computations such as convolution
and activation and the objects that store tensors such as parameters and gradients.
The class includes one method for the training loop and one method for testing.

The generated code includes the corresponding IR expressions in the comments 
to improve readability. For example, the code below shows the Java statements generated for 
the forward inference of max pooling. Note that the variable name in the comments has no relation to the variable names in the code as they are independently generated. 

\begin{lstlisting}[language=Scala]
  // val X9 = Pooling(2,2,0,true)(X8)
  JCudaTensor x16;
  JCudaTensor x17;
  x17 = x9;
  x16 = x18.forward(x17);
\end{lstlisting}

It is easy to perform some customization of the generated code such as 
changing the number of training iterations or reducing learning rate at specified interval. 
User can also use the generated code as a component of another application.

\paragraph{Persistency}
The compiled Java source includes code to save the trained parameters into files after
the training is complete. When the same program or another program compiled from the same network
starts, it can load the same parameters to resume training or for forward testing. 

\paragraph{Workspace}
The convolution layers in the compiled Java source share the same workspace. Thus, users
can place a limit on the total workspace by making one change.
By reducing workspace and using memory efficient mode, users may reduce memory use to fit into a particular GPU. 

\section{Performance}
\label{perform}

The primary compilation target of DeepDSL is Java program that runs on Nvidia GPU through its CUDA/CuDNN library\footnote{DeepDSL has limited support for CPU with features sufficient for implementing Lenet.}.
DeepDSL can encode well-known networks such as Alexnet, Overfeat, GoogleNet, Vgg, and Deep Residual Networks (Resnet).
In this section, we evaluate the performance of DeepDSL with Caffe and Tensorflow using these networks. 
To be consistent, DeepDSL, Caffe, and Tensorflow tests all follow the same Caffe prototxt definitions. 
Specifically,  for Alexnet and GoogleNet, we followed the prototxt from Caffe's website\footnote{\url{github.com/BVLC/caffe/tree/master/models}}; 
for Vgg (Vgg-16), we followed prototxt from this link\footnote{\url{github.com/ruimashita/caffe-train/blob/master/vgg.train_val.prototxt}};
for Overfeat, we followed prototxt from IntelLabs\footnote{\url{github.com/IntelLabs/Latte.jl/blob/master/benchmarks/overfeat/overfeat.prototxt}}; 
and for Deep Residual Network (ResNet-50), we followed the prototxt from the author's website\footnote{\url{github.com/KaimingHe/deep-residual-networks/blob/master/prototxt/ResNet-50-deploy.prototxt}}. 
The Tensorflow implementation of these networks are either modified from versions of convnet-benchmarks\footnote{\url{github.com/soumith/convnet-benchmarks}} or 
created from scratch. 
Note there is a couple of differences between the tests of Tensorflow and those of DeepDSL and Caffe. 
For example, the training data in the Tensorflow tests is generated from random data in memory 
while DeepDSL and Caffe tests load real images from the Lmdb database. 
In addition, the GoogleNet test of Tensorflow only includes the main branch of the GoogleNet while DeepDSL and Caffe train with the full network. 
All our tests are trained with ImageNet images that have been resized to 224 by 224 (though DeepDSL do supports random cropping of images when 
their sizes are larger than specified dimensions).

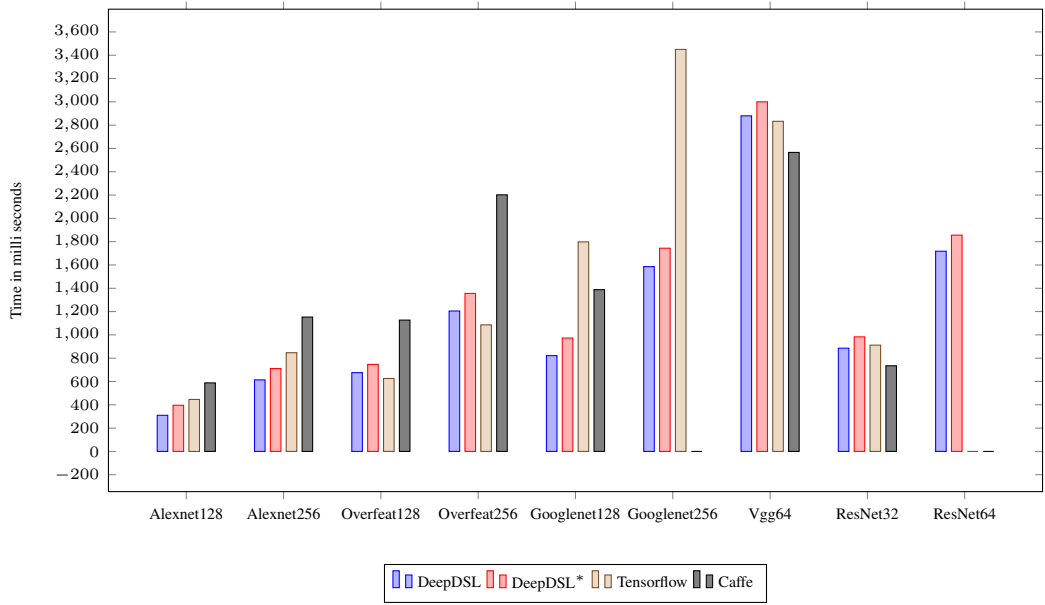
\begin{figure}[h]
\begin{tikzpicture}
\begin{axis}[
    tiny,
    height=8cm, width=14cm,
    ybar,
    enlargelimits=0.1,
    legend style={at={(0.5,-0.15)}, anchor=north,legend columns=-1},
    ylabel={Time in milli seconds},
    symbolic x coords={Alexnet128, Alexnet256, Overfeat128, Overfeat256, Googlenet128, Googlenet256, Vgg64, ResNet32, ResNet64},
    xtick=data,
    bar width = 4pt,
]
\addplot 
	coordinates {(Alexnet128, 310) (Alexnet256, 614) (Overfeat128, 676) (Overfeat256, 1205) (Googlenet128, 822) (Googlenet256, 1586) (Vgg64, 2880) (ResNet32, 886) (ResNet64, 1718)};

\addplot 
	coordinates {(Alexnet128, 396) (Alexnet256, 711) (Overfeat128, 747) (Overfeat256, 1356) (Googlenet128, 973) (Googlenet256, 1744) (Vgg64, 3000) (ResNet32, 983) (ResNet64, 1856)};

\addplot 
	coordinates {(Alexnet128, 446) (Alexnet256, 847) (Overfeat128, 626) (Overfeat256, 1086) (Googlenet128, 1798) (Googlenet256, 3450) (Vgg64, 2833) (ResNet32, 912) (ResNet64, 0)};

\addplot 
	coordinates {(Alexnet128, 588) (Alexnet256, 1153) (Overfeat128, 1127) (Overfeat256, 2202) (Googlenet128, 1388) (Googlenet256, 0) (Vgg64, 2566) (ResNet32, 735) (ResNet64, 0)};

\legend{DeepDSL, DeepDSL$^*$, Tensorflow, Caffe}
\end{axis}
\end{tikzpicture}

\caption{Runtime performance of DeepDSL, Tensorflow, and Caffe (1 forward/backward iteration). 
DeepDSL and DeepDSL$^*$ are performance in runtime-efficient and memory-efficient mode respectively. 
The names of the networks are followed by the batch size. 
Caffe failed to run GoogleNet (batch 256) and ResNet (batch 64) and 
Tensorflow failed to run ResNet (batch 64) due to the exhaustion of GPU memory.
}
\label{fig:runtime}
\end{figure}

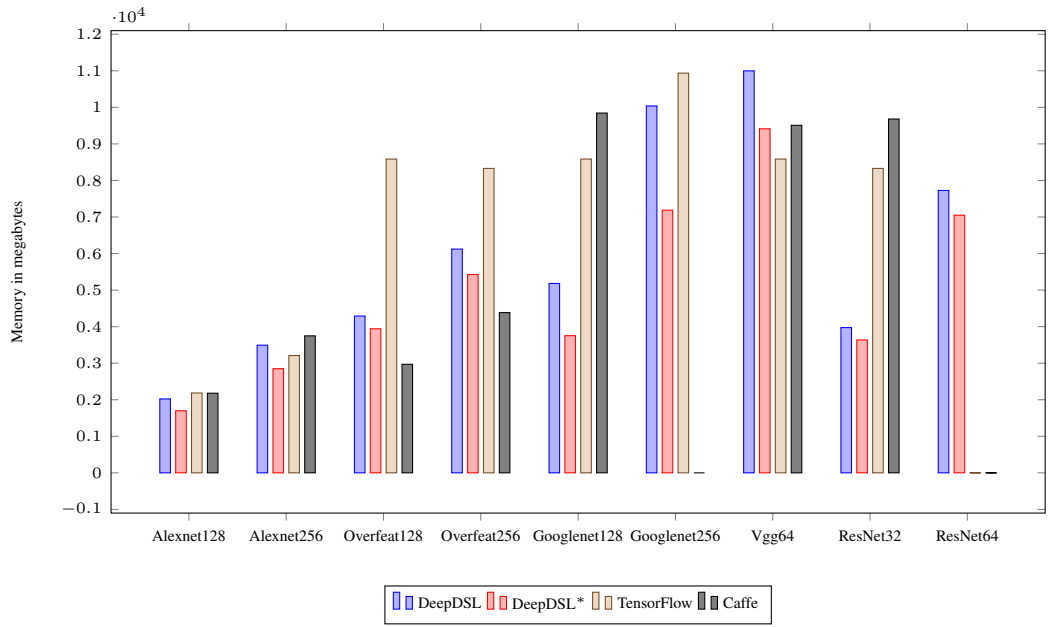
\begin{figure}[h]
\begin{tikzpicture}
\begin{axis}[
    tiny,
    height=8cm, width=14cm,
    ybar,
    enlargelimits=0.1,
    legend style={at={(0.5,-0.15)}, anchor=north,legend columns=-1},
    ylabel={Memory in megabytes},
    symbolic x coords={Alexnet128, Alexnet256, Overfeat128, Overfeat256, Googlenet128, Googlenet256, Vgg64, ResNet32, ResNet64},
    xtick=data,
    bar width = 4pt,
]
\addplot 
	coordinates {(Alexnet128, 2022) (Alexnet256, 3493) (Overfeat128, 4290) (Overfeat256, 6122) (Googlenet128, 5182) (Googlenet256, 10038) (Vgg64, 10998) (ResNet32, 3974) (ResNet64, 7726)};

\addplot 
	coordinates {(Alexnet128, 1698) (Alexnet256, 2849) (Overfeat128, 3941) (Overfeat256, 5425) (Googlenet128, 3752) (Googlenet256, 7187) (Vgg64, 9413) (ResNet32, 3636) (ResNet64, 7049)};

\addplot 
	coordinates {(Alexnet128, 2185) (Alexnet256, 3209) (Overfeat128, 8585) (Overfeat256, 8329) (Googlenet128, 8585) (Googlenet256, 10936) (Vgg64, 8585) (ResNet32, 8329) (ResNet64, -1)};

\addplot 
	coordinates {(Alexnet128, 2179) (Alexnet256, 3748) (Overfeat128, 2970) (Overfeat256, 4384) (Googlenet128, 9844) (Googlenet256, 0) (Vgg64, 9509) (ResNet32, 9681) (ResNet64, -1)};

\legend{DeepDSL, DeepDSL$^*$, TensorFlow, Caffe}
\end{axis}
\end{tikzpicture}
\caption{Peak GPU memory use of DeepDSL, Tensorflow, and Caffe during training.
DeepDSL and DeepDSL$^*$ are performance in runtime-efficient and memory-efficient mode respectively. 
Caffe ran out of GPU memory for Googlenet (batch 256) and ResNet (batch 64).
Tensorflow ran out of memory for ResNet (batch 64).
}
\label{fig:memory}
\end{figure}

The tests are run on a server with a single NVIDIA Tesla K40c GPU equipped with 12 gigabytes of memory. 
The server runs CentOS 7 Linux distribution. 
DeepDSL uses the JCuda 0.8.0RC binding that runs against CUDA 8.0.27\footnote{Note previous CUDA versions such as 6.5 or 7.x can also be used.}. 
DeepDSL programs are publicly available\footnote{\url{github.com/deepdsl/deepdsl}}. 

The runtime performance of DeepDSL, Tensorflow, and Caffe is compared in Figure~\ref{fig:runtime}, where
DeepDSL has significant advantage over Caffe in Alexnet, Overfeat, and Googlenet while only marginally slower than Caffe
in Vgg and ResNet (Deep Residual Network). 
DeepDSL is also faster than Tensorflow in Alexnet, Googlenet, and ResNet while slightly slower in Overfeat and Vgg. 

The memory consumption of the DeepDSL, Tensorflow, and Caffe is compared in Figure~\ref{fig:memory}, where DeepDSL uses less
memory in Alexnet, Googlenet, and ResNet while Caffe uses less memory in Overfeat and Vgg. 
DeepDSL uses significantly less memory for Googlenet and ResNet where Caffe runs out of memory for Googlenet at batch size 256 and ResNet at batch size 64. 
DeepDSL uses less memory than Tensorflow in all tests except Vgg. Tensorflow also ran out of memory for ResNet at batch size 64.
It is unclear why Tensorflow uses similar amount of memory for Overfeat with batch size 128 and 256.

In the tests, DeepDSL programs are run with runtime efficient mode which caches tensor objects and 
with memory efficient mode (denoted by DeepDSL$^*$) which deallocates tensor objects as soon as possible.
DeepDSL$^*$ uses 10 to 30\% less memory with similar percentage of runtime overhead 
except Vgg and Googlenet where runtime overhead is relatively smaller than memory saving.

DeepDSL also lets CUDNN to pick the convolution algorithms with max performance. 
In Overfeat (batch size 128), out of the 4290 megabytes of GPU memory consumed, more than 2700 megabytes are for convolution workspace. 
While Caffe uses less memory in this test, it also runs much slower.

Among all tests, DeepDSL either outperforms Caffe by a large margin or uses significantly less memory with Vgg being
the only exception where Caffe uses slightly less time and memory. DeepDSL also has competitive runtime performance when compared with Tensorflow. 

As a side note, while running DeepDSL requires little setup, installing libraries such as Caffe and Tensorflow requires
a list of dependencies and long compilation sessions. Consequently, we skipped testing with Torch 7 after a few failed 
attempts due to time limitation.

\section{Related work}
\label{rel}

In this section, we review some popular tools: Torch7, Theano, Caffe, TensorFlow, and CNTK, and 
newer ones such as Chainer~\cite{chainer_learningsys2015} and MXNet~\cite{chen2015mxnet}. 

Torch7~\cite{collobert:2011c} uses Lua language for integration with C program and achieves C-like performance. 
It has a large set of optimized routines to support CPU, GPU, mobile and FPGA backends.
Theano~\cite{bergstra+al:2010-scipy}, hosted in Python, 
allows users to define symbolic variables and functions (using NumPy~\cite{DBLP:journals/corr/abs-1102-1523}) 
to encode DL networks and compiles the symbolic expressions to C. 
Theano performs optimization such as normalizing mathematical expressions, numerical stabilization, and code specialization 
during compilation and the target code can run on CPU or GPU devices. 
Caffe~\cite{jia2014caffe} constructs a graph for DL network by connecting the layers with the 4D arrays that store tensors. 
Caffe separates its DL network model representation (using ProtocolBuffers~\cite{protobuf}) from the actual model parameter calculations. 
With its layered structure, Caffe computes the memory needed for each layer and reserves memory accordingly. 
TensorFlow~\cite{2016arXiv160304467A} shares largely common design paradigms as that of Caffe. Its core is written in C++ and 
its computation graph is described with a graph where tensors and layers are alternatively arranged. 
Unlike Caffe's tensor, TensorFlow's tensor is a typed multi-dimensional array and is persistent mutable. 
Like TensorFlow and Caffe, CNTK describes a network with a configuration file. 
CNTK can encode arbitrary computational network and it can map computation onto multiple GPUs across multiple machines 
by assigning each computation node to a particular CPU/GPU device. 

Comparing to the ``define-and-run'' paradigm (adopted by Torch7, Theano, and Caffe), 
Chainer~\cite{chainer_learningsys2015} follows a ``define-by-run'' pattern,
which essentially allows modifying the control flow during the execution of a computational graph. 
MXNet~\cite{chen2015mxnet} provides both declarative and imperative programming styles and 
multiple language supports by embedding into multiple host languages and unifying the execution with one backend engine.

The major difference between DeepDSL and the above tools is that DeepDSL is fully abstract until code generation. 
This means that DeepDSL's intermediate representation 
can be compiled to different languages or to run on different platforms. 
While the current compilation target of DeepDSL is Java, 
targeting a different language mainly involves building an interface library
to call CUDA routines while the optimization components of DeepDSL remain the same.
This separation between optimization and code generation also means that 
we can apply generic optimization techniques at IR level without worrying about the underlying data structure 
such as the representation of tensors or how the layers are connected.
In fact, the optimization of DeepDSL involves nothing specific to deep neural networks 
since they are mostly compilation techniques.

Note that while Theano and DeepDSL have similarity in the way that DSL expressions are optimized and transformed, 
there are two important differences that make DeepDSL more efficient and flexible. 

One is that while Theano expressions are treated as graphs during optimization,
DeepDSL expressions are optimized in two phases. 
The first phase is at expression level where the training loss and the parameter gradients
go through the process of simplification, loop merging, code motion, and vectorization. 
In the second phase, DeepDSL expressions are reduced to {\em static single assignment} form
for additional optimization such as common subexpression elimination, code scheduling, 
inlining of in-place computation, and tensor deallocation.

Two is that DeepDSL generates target code using a single-pass generator (about 1200 lines of code) 
that prints Java source code as strings to a file. The input of the generator is DeepDSL expressions, 
which are completely independent from the generated code. 
The generated Java code is high-level and human readable with a simple Java API that allows customization.
This clean separation between DSL expression and target code also allows independent evolution of 
DSL optimization and target-code generation.
In contrast, the code generation of Theano is embedded in its functions for low-level computation 
and is tied to C code that is not readable to users.

\section{Conclusion}
\label{conclude}

We have developed a domain specific language DeepDSL that compiles to Java source program for deep learning.
The compiled DeepDSL programs are very easy to use and extend as its primary dependencies are just JCuda and CUDA libraries. 
DeepDSL programs are also efficient and its runtime performance and memory consumption are 
significantly better than Caffe and Tensorflow in some DL networks. 
DeepDSL performs static analysis for early error detection and provides readable intermediate representation and memory consumption analysis. 
DeepDSL allows compact encoding of complex networks and since it is based on a strongly typed language Scala, writing DeepDSL programs is
less error prone than dynamic languages such as Python.

While the compiled DeepDSL programs are efficient, DeepDSL itself is not optimized. Though compiling simpler networks such as Alexnet takes a few seconds, the compilation of complex networks such as ResNet can take a few minutes. As the future work, we plan to optimize DeepDSL to improve the compilation efficiency. 
Also while the memory efficient mode of DeepDSL can reduce GPU memory consumption, it may not be enough for memory
intensive networks such as Vgg. As future work, we plan to implement GPU memory virtualization by paging out tensors that
are not immediately needed. 

\bibliography{doc}
\bibliographystyle{preprint}

\end{document}